\documentclass[10pt,conference]{IEEEtran}
\IEEEoverridecommandlockouts 

\usepackage{cite}
\usepackage{fontawesome5}
\usepackage{amsmath,amssymb,amsfonts}
\usepackage{algorithmic}
\usepackage{graphicx}
\usepackage{textcomp}

\usepackage[table,dvipsnames]{xcolor}

\usepackage[most]{tcolorbox}
\usepackage{tabularray}

\usepackage{todonotes}

\newcommand{\rqAUm}{How do LGBTQIA+ developers get involved in agile teams?}

\newcommand{\rqADois}{What are the expectations and challenges of LGBTQIA+ developers in agile teams?}

\newcommand{\rqATres}{What are the recommendations for improving the DX of LGBTQIA+ developers in agile teams?}
    
\begin{document}

\title{The Developer Experience of LGBTQIA+ People in Agile Teams: a Multivocal Literature Review\\
\thanks{This study was financed in part by the Coordenação de Aperfeiçoamento de Pessoal de Nível Superior - Brasil (CAPES) - Finance Code 001}
}

\author{\IEEEauthorblockN{Edvaldo R. Wassouf Junior}
\IEEEauthorblockA{\textit{FACOM} \\
\textit{UFMS}\\
Campo Grande, Brazil \\ edvaldo.junior@ufms.br
}
\and
\IEEEauthorblockN{Débora Paiva}
\IEEEauthorblockA{\textit{FACOM} \\
\textit{UFMS}\\
Campo Grande, Brazil \\
debora.paiva@ufms.br}
\and
\IEEEauthorblockN{Kiev Gama}
\IEEEauthorblockA{\textit{CIn} \\
\textit{UFPE}\\
Recife, Brazil \\
kiev@cin.ufpe.br}
\and
\IEEEauthorblockN{Awdren Fontão}
\IEEEauthorblockA{\textit{FACOM} \\
\textit{UFMS}\\
Campo Grande, Brazil \\ awdren.fontao@ufms.br
}
}

\maketitle

\begin{abstract}
Research on underrepresented populations is essential for fostering greater diversity within the software industry. Team diversity is important for reasons that go beyond ethics. Diversity contributes to greater innovation and productivity, helping decrease turnover rates and reduce team conflicts. Within this context, LGBTQIA+ software engineering professionals face unique challenges, e.g., self-isolation and invisibility feeling. Developer Experience (DX) encompasses cognitive, emotional, and motivational considerations, supporting the idea that improving how DX can enhance team performance, strengthen collaboration, and lead to more successful software projects. This study aimed to examine traditional and grey literature data through a Multivocal Literature Review focused on the DX of LGBTQIA+ professionals in agile teams. Our findings reveal that issues such as invisibility, prejudice, and discrimination adversely affect their experiences, compounded by the predominance of heterosexual males in the field. Conversely, professionals who feel welcomed by their teams and organizations, especially in processes tailored to their needs, report more positive team dynamics and engagement. 
\end{abstract}

\begin{IEEEkeywords}
  LGBT, LGBTQIA+, diversity, developer experience, agile.

\end{IEEEkeywords}

\section{Introduction}


Developer Experience (DX) refers to how developers perceive and feel about their activities, tools, and environments within software development. It emphasizes cognitive, emotional, and motivational factors, highlighting that improving DX positively impacts team performance, collaboration, and software project outcome~\cite{fagerholm2012developer}. The emotional factors of DX concern how developers feel about their work, encompassing aspects such as respect, belonging, attachment, and social and team dynamics, which are deeply connected to psychological safety \cite{fagerholm2012developer, razzaq2024systematic, greiler2022actionable}. For underrepresented groups, including LGBTQIA+ professionals, these factors are particularly critical, as they often face challenges in fostering a sense of belonging and inclusion in environments traditionally dominated by male and heterosexual norms~\cite{de2023benefits,de2023transgender, de2024hidden}.

In the context of agile software, the composition of teams and the dynamics of engagement among professionals are key to determining the success of delivering value to customers \cite{matsubara2023moving, tiwari2024great,zahl2023teamwork, santos2024software}. Thus, it is essential to keep these professionals engaged and to prevent team disintegration, which can lead to a loss of retention\cite{de2023post}. Retained developers are motivated to engage in transactions continuously and are willing to continue their relationship with their teams \cite{fontaoDevGo}.  While fostering LGBTQIA+ inclusion contributes to innovation and productivity, it is primarily a fundamental ethical commitment to ensuring dignity, equality, and belonging for all individuals, regardless of their identity. This research underscores the intrinsic value of inclusion as a cornerstone for both equitable workplaces and broader societal progress.


Gender studies focusing on women and the presence of gender bias in the workplace are gaining traction~\cite{trinkenreich2022women, petrescu2023women}, with researchers increasingly examining the benefits and difficulties associated with team composition that includes diverse cultures, ethnicities, and nationalities \cite{mason2024diversity}. However, there is still a lack of data regarding the DX of professionals from underrepresented groups, particularly those in the LGBTQIA+ community, within the agile industry \cite{de2023benefits, de2024hidden}. Furthermore, social dimensions, satisfaction, respect, psychological safety, and trust in team dynamics and agile ceremonies are essential to assess the DX~\cite{alami2023antecedents, fagerholm2012developer,ahmad2023psychological}.

Researchers have investigated how various work models affect different LGBTQIA+ populations \cite{ford2019remote, de2023benefits}.  Gender bias impacts not only women but also LGBTQIA+ professionals, including transgender women, non-binary and queer people \cite{prado2020trans, nicholson2022remote}. It is essential to understand the particular challenges faced by those professionals in the agile industry and to identify recommendations that can enhance their experiences.
By examining the diversity within work environments, we can better understand how it shapes DX, including the benefits of fostering inclusion, its impact on engagement, and the practical application of diversity policies in the industry \cite{welsch2024navigating, de2023benefits}. However, factors such as sexual prejudice, discrimination, and various forms of violence can detrimentally affect DX, as they undermine performance, motivation, and the sense of belonging for LGBTQIA+ developers \cite{de2023post, de2023benefits, de2023transgender}.

As Software Engineering (SE) is a practitioner-oriented field, the role of ``grey'' (i.e., non-academic) literature should be formally recognized and included in research~\cite{garousi2019guidelines}. In this sense, alongside peer-reviewed studies, materials produced directly by industry professionals can be highly valuable \cite{garousi2019guidelines, garousi2020benefitting}. Such resources reflect the daily experiences and insights of professionals within their communities. We performed a multivocal literature review to gather insights from academic and industry sources about the challenges faced by LGBTQIA+ professionals from a DX perspective. The aim is to synthesize recommendations to improve the DX of this underrepresented group. Key findings show that measures should prioritize the well-being of LGBTQIA+ professionals by tackling invisibility and discrimination in teams. Results highlight the need for effective inclusion policies for this underrepresented group.

\section{Background and Related Work}


\subsection{Diversity in Software Engineering}
In traditional literature, gender diversity in Software Engineering (SE) is frequently studied under a binary male-female perspective that leaves behind problems faced by individuals of various gender identities from the LGBTQIA+ groups. In a systematic mapping study about diversity in SE conducted by Silveira et al. \cite{silveira2019systematic}, the authors found 129 papers on gender identity concerning women and only two about LGBTQI. In the most recent literature review on diversity in SE~\cite{rodriguez2021perceived}, while there were 80 studies about gender diversity identified, there were only two focusing on transgender software engineers~\cite{ford2019remote,prado2020trans} and no mentions of other LGBTQIA+ studies.  Within a variety of sexual and gender identities in the LGBTQIA+ community, transgender professionals often face underrepresentation and invisibility~\cite{frluckaj2024paradoxes,de2024hidden}. A pioneering study examining the role of remote work in enabling the inclusion of transgender professionals was conducted by Ford et al.\cite{ford2019remote}. The authors argue the importance of controlling identity disclosure and promoting safe disengagement from harmful interactions. In addition, they suggest further research to improve support for marginalized groups in technology through remote practices. Prado et al.\cite{prado2020trans} analyzed the inclusion of transgender professionals in hackathons, identifying challenges such as discrimination and identity invalidation. 

Overall, the effects of workplace discrimination and prejudice on the productivity and well-being of LGBTQIA+ developers, which are aspects that can negatively impact the developer experience, remain underexplored in the literature. Recent research on workplace discrimination in SE conducted by Zhao et al. \cite{zhao2023workplace} aimed to understand this issue through the lens of male and female genders, however, there is a gap in the experience of professionals who identify beyond these genders that can be addressed in studies with this focus. Studies addressing this aspect are limited, such as the study conducted by e Souza and Gama \cite{de2020diversity}, where the respondent comment the minimization of episodes of lgbtphobia against LGBTQIA+ developers. de Souza Santos et al. \cite{de2023benefits} discussed how remote work improves psychological safety for LGBTQIA+ professionals, reducing exposure to discrimination and allowing identity control. Still, it runs the risk of isolation without inclusive practices. 



\subsection{Developer Experience (DX)}

Fagerholm and Munch \cite{fagerholm2012developer} when defining Developer Experience (DX), reiterate that experience does not refer to expertise, but rather to the involvement of developers in software development activities. In a conceptual framework defined by them, three axes delimit DX: Cognition (techniques, platform, process, skill, procedures), Conation (Plans, goals, alignment, commitment, motivation, intention), and Affect (respect, team, social, attachment, belonging). In this sense, DX becomes a lens for analyzing and improving the experience, observing the main factors that influence productivity, engagement, and job satisfaction \cite{fagerholm2012developer}, \cite{greiler2022actionable}.

The tech industry uses strategies such as surveys to verify satisfaction and productivity to improve the DX of professionals. Many of these processes are shared in the field as in the research conducted at Google by D’Angelo et al. \cite{d2024measuring}. DX is often analyzed from a very technical perspective around tools and technologies, but it is a highly personalized experience, varying significantly between individuals. DX is shaped by a combination of individual, organizational, and technical challenges. Concerning the \textit{affect} dimension, the concept of psychological safety is a key driver of team and business performance and is critical in DX as noted by Greiler et al.\cite{greiler2022actionable}.

The concept of psychological safety is a shared belief among team members that they can take interpersonal risks, such as expressing ideas, admitting mistakes, or seeking help, without fear of criticism or blame \cite{alami2023antecedents}.  Furthermore, the concepts of psychological distress and psychological safety were captured as factors influencing DX in the Systematic Literature Review performed by Razzaq et al. \cite{razzaq2024systematic}. The authors identified 33 factors influencing DX and distilled them into 10 core themes. When reviewing the synthesis of knowledge obtained through the review, a gap is noted in the literature on DX for underrepresented groups in the software industry - this includes LGBTQIA+ professionals who face unique challenges in the face of prejudice and discrimination in software development.

In software teams, it is crucial to foster open communication, collaboration, and innovation, enabling teams to effectively address challenges and improve performance \cite{greiler2022actionable, de2023benefits}. In addition, although traditional literature was thoroughly reviewed, data from industry professionals in grey literature sources (e.g., forums and online posts) were not incorporated. This grey literature data is essential~\cite{garousi2020benefitting} for understanding the lived experiences of these professionals and enhancing their workplace experience.

Our research focused on following recommendations from diversity researchers in the technology sector to explore the experiences of LGBTQIA+ professionals in the software industry \cite{de2023benefits, de2020diversity, ford2019remote, de2024hidden}. We gathered evidence by examining peer-reviewed literature and analyzing materials shared by LGBTQIA+ professionals in the industry. Our goal is to synthesize recommendations to improve the experience of this underrepresented group.
\section{Method}


\subsection{Goal and Research Questions}
We used the GQM \cite{basili1994gqm} to define our goal and RQs. Our goal is to \textbf{analyze} scientific evidence and material produced by SE practitioners outside academic forums \textbf{aiming to} understanding challenges and expectations, \textbf{concerning} developer experience \textbf{from the viewpoint of} researchers and LGBTQIA+ developers \textbf{in the context of} agile teams.  The main research question (RQ) for this multivocal systematic literature review is: \textbf{What do the scientific literature and grey literature say about the DX of LGBTQIA+ developers in agile teams?} To help answer this the following sub-questions were used:

\textbf{A.1: \rqAUm} \textbf{\textit{ Rationale:}} it explores how team integration happens in a field dominated by heterosexual male norms.

\textbf{A.2: \rqADois} \textbf{\textit{ Rationale:}} it aims to uncover how their experiences align or diverge from their expectations.

\textbf{A.3: \rqATres} \textbf{\textit{ Rationale:}} it seeks to synthesize evidence-based recommendations to create environments where LGBTQIA+ developers can thrive. 

\subsection{Multivocal Systematic Literature Review}
To address the research question, we used the multivocal systematic literature review method that has been widely used in SE  research because it provides methodologies to categorize published studies and in areas where primary studies are scarce or not very relevant \cite{garousi2020benefitting,garousi2016need}. 
Following systematic review guidelines from Garousi et al. \cite{garousi2019guidelines}, which advocate for the inclusion of grey literature to broaden perspectives and insights, we employed these strategies to formulate research questions and execute a multivocal systematic literature review. The planning and execution of our multivocal literature review were organized into five distinct stages:


\noindent\textit{- Stage 1)} We began by defining our research questions and composing a search string. This string was refined multiple times and control studies were identified. We reviewed the scope and continued to refine the string until we achieved saturation, which we tested using the IEEExplore platform. 

\noindent\textit{- Stage 2)} Next, we defined our inclusion and exclusion criteria. We executed the search\footnote{https://bit.ly/amcSearch  https://bit.ly/ieeeSearch  Our access to the Scopus platform generated a link via the institution's interface.} on various databases, including IEEExplore, ACM Digital Library, and Scopus Elsevier. 

\noindent\textit{- Stage 3)} We classified the studies and updated the systematic review employing backward and forward snowballing. Then, we conducted a thematic synthesis of all included studies.

\noindent\textit{- Stage 4)} We examined the use of grey literature.  The use of grey literature can be an interesting mechanism to fill the gap in the connection with the practice of SE (Software Engineering) and by people who research and work with SE in academic contexts \cite{garousi2020benefitting}. Inclusion and exclusion criteria were established, and we constructed two different search strings. The first string was used to index documents directly from the Dev\footnote{https://dev.to/} platform, while the second string utilized the \textit{Google Search Engine} to index documents from the same platform. Both strings were refined before the searches were executed. 

\noindent\textit{- Stage 5)} Finally, we classified the returned documents and performed a thematic synthesis of the documents and posts.

\section{Systematic Literature Review}
We performed our systematic review\footnote{ keyword extraction, systematic review, snowballing https://figshare.com/s/04c5471717429e121110} (SLR) according to the guidelines proposed by Kitchenham et al. \cite{kitchenham2015evidence}. Our SLR was conducted from March to May 2024, covering publications up to 2023 across various databases. Our protocol outlines the planned procedures for conducting the systematic review, covering the search strategy, study selection, data extraction, and data analysis. Additionally, it clarifies the primary responsibilities of each co-author. The first author prepared the initial draft of the protocol, which was then reviewed by all authors.

\subsection{SLR Method}

\noindent\textbf{QASI (Quasi-Gold Standard):} Control studies (QASI) were used to verify the relevance of database search results after composing the search string \cite{zhang2011empirical}. The selected control studies provide key terms related to gender\cite{kohl2018perceptions} and sexual diversity\cite{de2023benefits} within the context of software development, as well as minority groups related to sexual diversity. \textbf{Table \ref{QASI}} presents these selected works.

\begin{table}[!ht]\footnotesize
\centering
\fontsize{6.9pt}{6.9pt}\selectfont

\caption{QASI: Quasi-Gold Standard}
\begin{tabular}{|l|l|l|} 
\hline
\textbf{ID} & \multicolumn{1}{c|}{\textbf{Authors}} & \multicolumn{1}{c|}{\textbf{Title}} \\ 
\hline
C1 & \begin{tabular}[c]{@{}l@{}}Kohl, Karina; \\Prikladnicki, Rafael;\end{tabular} & \begin{tabular}[c]{@{}l@{}}Benefits and difficulties of gender\\diversity on software~development\\teams: A qualitative study.\end{tabular} \\ 
\hline
C2 & \begin{tabular}[c]{@{}l@{}}de Souza Santes, Ronnie;\\de Magalhaes,Cleyton;\\Ralph, Paul~\end{tabular} & \begin{tabular}[c]{@{}l@{}}Benefits and limitations of remote\\work to~LGBTQIA software\\professionals.\end{tabular} \\ 
\hline
C3 & \begin{tabular}[c]{@{}l@{}}de Souza Santos, Ronnie; \\Stuart-Vermer, Brody;~\\de Magalhães, Cleyton;\end{tabular} & \begin{tabular}[c]{@{}l@{}}What do transgender software\\professionals~say about a career\\in the the software industry?\end{tabular} \\ 
\hline
C4 & \begin{tabular}[c]{@{}l@{}}de Souza Santos, Ronnie;\\Stuart-Vermer, Brody;\\de Magalhães, Cleyton;\end{tabular} & \begin{tabular}[c]{@{}l@{}}Diversity in software engineering:\\A survey about scientists from \\underrepresented groups.\end{tabular} \\ 
\hline
C5 & \begin{tabular}[c]{@{}l@{}}de Souza Santos, Ronnie;\\Stuart-Vermer, Brody;~\\de Magalhães, Cleyton;\end{tabular} & \begin{tabular}[c]{@{}l@{}}LGBTQIA (In) Visibility in \\Computer Science~and \\Software Engineering Education.\end{tabular} \\ 
\hline
C6 & \begin{tabular}[c]{@{}l@{}}Blincoe, Kelly;\\Springer, Olga;\\Wrobel, Michal;\end{tabular} & \begin{tabular}[c]{@{}l@{}}Perceptions of gender diversity's\\impact on~mood in software\\development teams.\end{tabular} \\
\hline
\end{tabular}
\label{QASI}
\end{table}

\noindent\textbf{Search String:} We used the PICO strategy \cite{petersen2015guidelines} to frame relevant theoretical references to build the search string. The Population considers terms to the LGBTQIA+ population. In the intervention, terms from the domain of methodologies and agile teams were added, in addition to terms related to software development. A final string (Table \ref{string}) was obtained that returned the control works and the searches became saturated, no longer returning works related to the target theme.

\begin{table}[ht!]

\caption{String used in the Systematic Literature Review}
\centering
\begin{tblr}{
  width = \linewidth,
  colspec = {Q[940]},
  hlines,
}
{((``lgbt*'' OR ``lgbtqia+'' OR  ``gender diversity'' OR ``underrepresented groups''\\OR ``sexual diversity'' OR ``trans'' OR ``queer'' OR ``non binary'') AND \\(``software development'' OR ``agile development'' OR ``software professionals''\\~OR ``mobile team'' OR ``devops'' OR ``remote team'' OR ``software engineering''\\OR ``agile remote team'' OR ``agile software development teams'' OR ``squad''\\~OR ``tribe'' OR ``scrum'' OR ``xp''))}
\end{tblr}
\label{string}
\end{table}

\begin{figure}[ht!]
    \centering
    \includegraphics[width=1.05\linewidth]{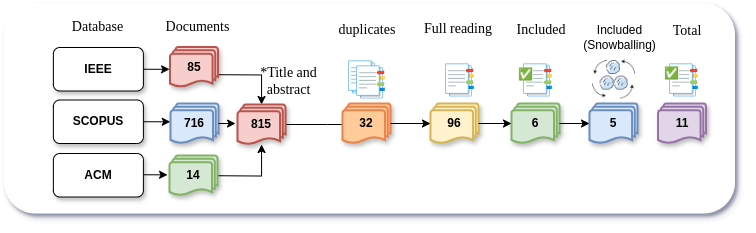}
    \caption{Process of filtering, classifying and analyzing articles from review and snowballing.}
    \label{fig:filtering}
\end{figure}

\noindent\textbf{Studies Selection:} Figure \ref{fig:filtering} illustrates the search performed in three databases. It yielded 815 studies distributed across the ACM Digital Library, IEEE, and Scopus. We extracted the authors' data, title, journal/conference information, and the DOI for each paper. After constructing the spreadsheet, the paper filtering criteria were defined, following the model used in~\cite{fontaoDevGo}. We applied the following inclusion/exclusion criteria: (A) Peer-reviewed studies; (B) Studies written in English; (c) Studies that answer at least one auxiliary question; (D) Study available on the Web or by contacting the authors; (E) Non-duplicated studies. Studies that met all the criteria were included. Otherwise, the study was excluded.

\begin{figure}[ht!]
    \centering
    \includegraphics[width=0.9\linewidth]{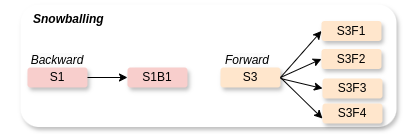}
    \caption{Snowballing performed to enrich the literature review}
    \label{fig:snowballing}
\end{figure}

\noindent\textbf{Snowballing:} We chose to enrich the systematic mapping of traditional literature with the \textit{Snowballing (Backward and Forward)} technique, guided by the review updates provided by \cite{garner2016and},\cite{felizardo2016using}. In this way, new works can be obtained by applying \textit{Snowballing} to the studies included in the mapping. The papers were classified into codes by iterating the \textit{Forward Snowballing } technique on the six included papers. The suffix 'F' was added to each included paper followed by an identifying code (numbered in order of discovery), which allowed the papers to be tracked.
The same process was adopted when applying the \textit{Backward Snowballing } technique, adding the suffix 'B' to the code of the included study. Four papers resulting from the application of \textit{Forward Snowballing } went through the criteria evaluation process and were included. Considering the four papers included, all of these resulted from the iteration on paper \textbf{S3}. Through the application of \textit{Backward Snowballing}, one study was included from study \textbf{S1}. There were no more papers included after applying the \textit{Snowballing} technique (Fig. \ref{fig:snowballing}). Then, the five articles included in the systematic mapping of traditional literature and the application of \textit{Snowballing} in the included works were listed with a unique tracking code. Table \ref{studies_included} lists all the works included in the mapping of traditional literature (six studies) enriched with \textit{Snowballing} (five studies), the \textbf{ID} "\textbf{S}" was added, according to the order of discovery.

\noindent\textbf{Quality Assessment:} We applied the quality criteria (QAC) following recommendations from Yang et al. \cite{yang2021quality}. The questions that supported the quality assessment are: Is the study relevant to Software Engineering? Is the purpose of the study clear? Does the study present the methodology used? Are the study’s contributions clear? Was the research published in a reputable venue?

\noindent\textbf{Data Analysis - Thematic Synthesis:} To search for themes on the findings of the selected studies, we applied the recommendations from Cruzes and Dyba \cite{cruzes2011recommended} to perform the thematic\footnote{https://figshare.com/s/eb209548db4f49731832} synthesis of the research. Our findings were synthesized by organizing codes into broad categories based on their labels, with each code assigned to a single category. Quotations linked to each code were analyzed in their original context, and observations were used to draft concise code descriptions. The categories were then refined iteratively, involving the reassignment of some codes and adjustments to category definitions, until a final structure was established. Any additional inconsistencies or errors in the extraction were identified and promptly addressed through direct communication with the original extractors. In total, we obtained 24 themes\footnote{Codebook SLR: https://figshare.com/s/6146d5c8595e26258b02} after processing the thematic synthesis. 

\begin{table}[ht!]
\fontsize{5pt}{5pt}\selectfont

\centering
\caption{Studies included}
\begin{tblr}{
  column{3} = {c},
  cell{1}{2} = {c},
  cell{1}{4} = {c},
  hlines,
  vlines,
}
\textbf{ID} & \textbf{Authors} & {\textbf{Research}\\\textbf{~Databases}} & \textbf{Title \& Year}\\
S1 & {Poncell, Igor;\\Gama, Kiev;} & ACM & {Diversity and Inclusion Initiatives \\in Brazilian Software~Development\\Companies: Comparing the \\Perspectives of~Managers and \\Developers. (2022)}\\
S2 & {de Souza Santos, Ronnie;\\de Magalhaes, Cleyton;} & ACM & {Benefits and limitations of remote\\work to LGBTQIA+~\\software professionals. (2023)}\\
S3 & {Ford, Denae;\\Milewicz, Reed;\\Serebrenik,~~Alexander ;} & ACM & {How remote work can foster a \\more~inclusive environment for\\~transgender developers. (2019)}\\
S4 & {de Souza Santos, Ronnie,\\Stuart-Verner, Brody;\\Magalhães, Cleyton  } & IEEE & {What do transgender software \\professionals say about a~\\career in the software industry. (2023)}\\
S5 & {Wang, Yi;~\\Xinyue Zhang;\\Wei Wang ;} & IEEE & {Fundamentalists, Integrationists,\\Transformationists:~An Empirical\\Theory~of Men Software~Engineers’\\Orientations~in Gender Inequalities. (2023)}\\
S6 & {Kohl, Karina;\\Musse, Raupp;\\Manssur, Isabel; \\Vieira, Renata;~\\Prikladinick, Rafael ; } & SCOPUS & {Reinforcing diversity company \\policies:~Insights from~Stackoverflow\\developers survey. (2019)}\\
S7 (S1B1) & {de Souza, Natália;~\\Gama, Kiev ;} & IEEE & {Diversity and inclusion: Culture and\\perception in~information~technology\\companies. (2020)}\\
S8 (S3F1) & {Prado, Rafa;~\\Mendes,~Wendy;\\Gama, Kiev;\\Pinto, Gustavo;} & IEEE & {How trans-inclusive are \\hackathons? (2020)}\\
S9 (S3F2)& {Gunawardena, Sanuri;~\\Devine, Peter;~~\\Beaumont, Isabelle;\\Garden, Lola;~\\Murphy-Hill, Emerson;~\\Blincoe, Kelly ; } & ACM & {Destructive criticism in software code\\review impacts~inclusion. (2022)}\\
S10 (S3F3)& {de Souza Santos, Ronnie;~\\Adisaputri, Gianisa; \\Ralph, Paul ;} & IEEE & {Post-pandemic Resilience of Hybrid \\Software Teams. (2023)}\\
S11 (S3F4) & {Popoola, Gabriel;~\\McKie, Morgan;~\\Moten, Jade; \\Fletcher, Trina ; } & IEEE & {Remote Work and Satisfaction for \\Black Engineers~and\\Computer Scientists. (2022)}
\end{tblr}
\label{studies_included}
\end{table}

\subsection{SLR Results}
\begin{tcolorbox}[colback=WildStrawberry!10!white,colframe=WildStrawberry!75!black,title=A.1: \rqAUm]
\footnotesize
\textit{Studies show that diverse teams often achieve higher productivity and resilience, leveraging unique perspectives in problem-solving (S1, S10). In remote work (S2), LGBTQIA+ professionals benefit from a sense of safety and control, which aids team integration and performance, though challenges like isolation and invisibility persist. This dynamic supports agile environments by fostering psychological safety while highlighting areas for improved visibility and inclusive communication.}
\end{tcolorbox}


To support the understanding of the themes identified in the studies, based on the thematic synthesis, themes associated with developers' perceptions about agile teams dynamics are marked with the symbol \faComments[regular].

In \textbf{S1}, the focus was on \textbf{\faComments[regular] \textit{low perceived D\&I }(Diversity and Inclusion)}. The authors interviewed managers and people from underrepresented groups, including LGBTQIA+ developers from three companies. From the perspective of some respondents, engagement in affirmative actions for inclusion was difficult to advertise because the communication tool (\textit{Slack}) was crowded with work-related messages. Still, they recognized the importance of these actions. In one of the companies, the diverse teams' productivity was better than the non-diverse teams. The study \textbf{S1} highlights the \textbf{\faComments[regular] benefits of diversity} when reporting higher performance from a diverse team compared to a less diverse team, aligning with existing literature. It also notes how the perception of a welcoming and diverse environment affects the performance of LGBTQIA+ software developers and other underrepresented groups. Similarly, \textbf{S10} (a survey on hybrid work and resilience) had 23\% of respondents identifying as LGBTQIA+. The authors emphasize that diverse teams bring varied perspectives and life experiences, enhancing problem-solving and overcoming challenges.

In \textbf{S2}, a dynamic of greater involvement between developers and the team is evidenced, in the \textbf{\faComments[regular] \textit{remote work environment}}; the authors investigated a sample of 57 software professionals, with different genders, ethnicities, and orientations, belonging to the LGBTQIA+ community. When interpreting the experiences of professionals, using \textit{Grounded theory methodology} (GTM) in \textbf{S2}, the authors report that remote work benefits LGBTQIA+ people, this is due to increased safety and visibility, which facilitates involvement with the team, since when integrated into a team, professionals who fear violence and discrimination feel safe, in addition, gains are made about control over their identity.

The authors in \textbf{S2} discuss the dynamics of team involvement, noting that individuals may choose to share their LGBTQIA+ identity later in remote work setups. However,
challenges such as \textbf{\faComments[regular] invisibility and isolation} for LGBTQIA+ professionals are reported as frustrations. This issue of invisibility in \textbf{S2} relates to team integration and engagement, impacting developers' motivation in agile environments. Another finding from \textbf{S2} is that LGBTQIA+ developers can adapt well to remote work because \textbf{\faComments[regular] they can control their environment} through camera settings and chat.  This control is beneficial, as it enhances psychological safety for this underrepresented group, which often faces discrimination and violence.

\begin{tcolorbox}[colback=Emerald!10!white,colframe=Emerald!75!black,title=A.2: \rqADois]
\footnotesize
\textit{LGBTQIA+ developers in agile teams expect a welcoming environment with representation and effective inclusion policies, viewing remote work positively due to increased safety and control over identity disclosure (S1, S2, S10). Challenges include dealing with discrimination, microaggressions, and balancing participation in D\&I initiatives with work demands, often exacerbated in in-person settings (S4, S7, S11). Control over identity sharing and protective policies are essential for psychological and physical safety in these agile environments (S8, S10).}
\end{tcolorbox}



To support the understanding of the themes identified in the studies, based on the thematic synthesis, themes associated with developers' expectations are marked with the symbol \faFlag[regular], while themes related to challenges are marked with \faFrown[regular].

In \textbf{S1}, the respondent expresses their \textbf{\faFlag[regular] expectations regarding representation}. The respondent states that a leadership figure from underrepresented groups helped her consider the professional journey. In addition, participants also report the \textbf{\faFrown[regular] challenge of participating in diversity and inclusion events} proposed by their employer, as they need to reconcile participation in events with work demands and a work tool (\textit{Slack}) with a large volume of information.

Study \textbf{S4} presents accounts from transgender participants working as software professionals. \textbf{\faFlag[regular] Expectations about team acceptance} are reported. In addition, there is \textbf{\faFrown[regular] frustration} about the presence of tokenism regarding invitations for trans people to participate in discussions about diversity in the company, which should be the responsibility of the company to promote a safe and inclusive environment.

Another point discussed in \textbf{S4} is the report of the transgender developer, about the importance of working remotely and the fact that she can control the camera on days when she is having dysphoria crises (a psychological process in which there is an aversion to one's own image). Still on transgender people, in study \textbf{S6}, it is reported that \textbf{\faFlag[regular] companies that have inclusion and diversity policies} are preferred by non-binary and transgender professionals participating in the research.

Study \textbf{S11} reports the experience of a \textit{\textbf{queer}} person and software professional, about their suffering with returning to in-person work, since they report a routine of micro-aggressions in the workplace and the loss of autonomy. There is also an \textbf{\faFlag[regular] expectation} about hybrid work, as opposed to returning to the in-person work model. This is consistent with the findings of \textbf{S2}, which describes the challenge of LGBTQIA+ professionals when dealing with \textbf{\faFrown[regular] discrimination}, possible aggression and violence in the in-person work model. In addition to possible \textbf{\faFrown[regular] toxic environments} faced in the in-person work by the research participants. This data is consistent with the reports brought in \textbf{S7} about the experiences of LGBTQIA+ professionals who suffered discrimination and sexism.

In work \textbf{S8}, the participants reported problems related to \textit{hackathons} organizations, events that are part of the agile scope. The research participants reported suffering \textbf{\faFrown[regular] discrimination} by the \textit{hackathons} organization and a transgender person also reported that she stopped participating in \textit{hackathons} due to discrimination suffered. \textbf{\faFrown[regular] Control over the disclosure of one's identity} was also a challenge presented by a trans person when participating in the \textit{hackathon}. This data is also reported in studies \textbf{S2, S3}. This fact is important for LGBTQIA+ developers, to express what information other people can access and when to share it \cite{ford2019remote, de2023benefits}.

\textbf{\faFrown[regular] Control over identity sharing} is extremely important, since by exposing one's gender identification, pronouns, or sexuality, the LGBTQIA+ community can be victims of discrimination and violence \cite{david2017capital, ford2019remote, de2023benefits, prado2020trans}. In \textbf{S2} this issue is also brought up and emphasized as essential for the safety and better experience of LGBTQIA+ people in agile software development environments.

\textbf{S10} highlights  \textbf{\faFlag[regular] practical actions}: creating a safe, inclusive environment for LGBTQIA+ \textbf{\faFlag[regular] individuals requires managing identity disclosure}. In the study \textbf{S9} contributes to this discussion, it only partially addresses the issue, as non-binary participants note that \textbf{\faFrown[regular] frustration about destructive feedback} in \textit{code reviews}. Despite its limited contribution, it's important to recognize the scarcity of studies on the experiences and challenges faced by non-binary individuals in agile software development.

In a study conducted in \textbf{S7} involving two companies with inclusion and diversity programs and policies, two respondents from each company were interviewed. One interviewee described experiencing rude and  \textbf{\faFrown[regular] discriminatory treatment} when interacting with the product owner of another team. Conversely, another participant noted that their current company, which implements \textbf{\faFlag[regular] D\&I policies}, offers better opportunities compared to their previous employer. The feelings of \textbf{\faFrown[regular] frustration and abandonment} identified by the authors stem from the \textbf{\faFrown[regular] inadequacy of the inclusion and diversity initiatives}. A participant in \textbf{S7} states that her leader minimizes incidents of homophobia within the team. A self-identified lesbian describes experiencing hypersexualization from her superior. The four participants express dissatisfaction with the effectiveness of D\&I policies, highlighting that such policies alone are insufficient for retaining LGBTQIA+ professionals.

In study \textbf{S5}, the research examines the mindset change of participants from fundamentalists regarding gender diversity to integrators and transformers in software development, \textbf{\faFlag[regular] promoting greater acceptance of gender diversity}. One participant noted that this \textbf{\faFrown[regular] transition was challenging}, as it coincided with their acceptance of their sexuality—a process unique to LGBTQIA+ individuals that cisgender heterosexual people do not experience.

When analyzing the contributions of the works, positive expectations regarding diversity in teams for LGBTQIA+ professionals are noted in \textbf{S1, S4, S7, S6, S10}. In studies \textbf{S2, S3, S8, S10}, the importance of controlling identity sharing/disclosure is noted. Challenges related to fear of physical and psychological violence, and discrimination/LGBTphobia suffered by professionals are reported in \textbf{S2, S4, S7, S8, S11}. Positive expectations about remote work are present in \textbf{S2, S3, S10, S11}.

In \textbf{S1} \textbf{S2}, \textbf{S7} \textbf{S8}, \textbf{S10} they address reflections on the importance of effective inclusion and diversity policies for LGBTQIA+ professionals. In \textbf{S2}, \textbf{S3}, \textbf{S7}, \textbf{S8}, \textbf{S11} points such as psychological safety, physical safety, LGBTphobia, and microaggressions are pointed out as challenges experienced by LGBTQIA+ professionals within their teams. The expectations of professionals for remote work as opposed to in-person work are found in \textbf{S2, S3, S4, S10, S11}.

\begin{tcolorbox}[colback=cyan!10!white,colframe=cyan!75!black,title=A.3: \rqATres]
\footnotesize
\textit{The studies recommend offering remote or hybrid work options, promoting democratic structures that allow workspace choice, and fostering inclusive recruitment and onboarding (S2, S9, S10). Establishing LGBTQIA+ committees and supportive networks helps reduce isolation, while inclusive hackathon organization and visible anti-discrimination policies can enhance participation and safety (S8, S10). Recognizing diversity as a driver of innovation also builds a welcoming culture, supporting retention and broader interest in tech fields.}
\end{tcolorbox}



Aiming to support the connection between recommendations found in literature and DX dimensions (affect, conation, cognition).\textbf{ We present below the identified recommendations for each DX dimension.}

Concerning \textbf{affect},  in \textbf{S2}, recommendations include promoting a culture of diversity and inclusion, creating committees for LGBTQIA+ developers to connect, celebrating diversity and inclusion, and recognizing diversity as a driver of technological development. In \textbf{S8}, recommendations for hackathon organization emphasize inclusive communication, increased participation of transgender individuals in organizing teams, and the establishment of codes of conduct against LGBTQ+phobia to ensure a welcoming and safe environment.

Regarding \textbf{conation}, in \textbf{S2}, the development of democratic remote work structures is recommended, allowing professionals to choose their workspace and helping LGBTQIA+ software developers address violence, toxicity, and challenges related to in-person work. \textbf{S10} advocates for fair recruitment practices, inclusive onboarding, and democratic remote work structures, highlighting the specific needs of LGBTQIA+ professionals and emphasizing diversity as a driver of innovation and investment. In \textbf{S8}, transparency in selection processes is recommended to facilitate access for trans individuals to job opportunities in the software industry, addressing challenges highlighted in \textbf{S6}.

Finally when analyzing \textbf{cognition}, \textbf{S9} suggests remote or hybrid work as a factor for improving the developer experience. However, while the study intersects with LGBTQIA+ issues, its primary focus is on racial matters, providing only partial insights into the question. \textbf{S2} addresses the benefits and limitations of remote work, noting a trend in 2024 towards in-person models. Gaps remain in addressing the in-person model, but challenges can be mitigated by offering remote work options.

\section{Grey Literature Review}
We followed the recommendations for systematic grey literature reviews from Garousi et al.\cite{garousi2019guidelines}\cite{garousi2020benefitting} and Kamei et al.~\cite{kamei2021evidence}. Many SE studies overlook community platforms like Dev.to\footnote{https://dev.to/} and Medium as data sources~\cite{liang2024controlled}. We selected the Dev.to database for its reputation as a prominent hub where software professionals share insights and discuss industry topics~\cite{papoutsoglou2021mining}. It was recently used as the single source of another GLR~\cite{cerqueira2024empathy} and in a mining study~\cite{papoutsoglou2021mining}. We excluded Medium due to the prevalence of paywalled articles, whereas Dev provides free access to content.

\subsection{GLR Method}
\label{glrMethod}
\noindent\textbf{Test for use of grey literature:} Due to the small sample of works found in the traditional literature to collaborate with the research questions, we performed the test proposed by Garousi et al. \cite{garousi2019guidelines} obtaining "Yes" in more than three questions. Thus, we were able to verify the use of literature in our research and design the search strategy.

\begin{figure}[ht]
    \centering
    \includegraphics[width=0.75\linewidth]{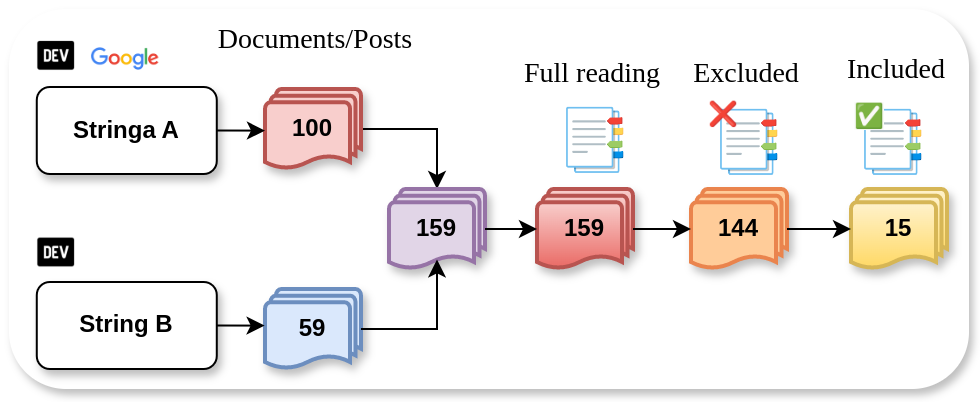}
    \caption{Grey literature review process/method execution.}
    \label{fig:process_grey}
\end{figure}

\noindent\textbf{Search Strategies:}
Two strategies were defined to search for material on the Dev.to. The first (Strategy A) consists of constructing a search string that is different from that used in SLR. In Strategy A, the search for content on dev.to was indexed by the Google Search Engine. The second (Strategy B) consists of searching directly on the Dev.to for terms related to the scope of this research. The GLR process is described in Fig. \ref{fig:process_grey} and the strings used in Table \ref{string_grey}.

To search for grey literature documents, the Chromium\footnote{https://www.chromium.org/} browser was used in anonymous mode. A stopping criteria (called Effort bounded \cite{garousi2019guidelines}) was defined to capture up to 100 documents/\textit{posts} for both search strategies based on the Google ranking algorithm. 
According to Google's search and indexing tools, to index results from a website, "site:address" must be added to the end of the search. Thus, to execute search strategy A, a search string was constructed using the PICO strategy \cite{petersen2015guidelines}, with the addition of site:dev.to. The search strategy A led to 100 posts according to the previously defined stopping criteria. In search strategy B, 59 documents referring to user posts were returned.

\begin{table}[ht!]

\caption{Search strategies used in the grey literature review}
\centering
\begin{tblr}{
  width = \linewidth,
  colspec = {Q[600]Q[340]},
  cell{1}{1} = {c},
  hlines,
  vlines,
}
\textbf{Strategy A} & \textbf{Strategy B}\\
{(``gay'' OR ``Bisexual'' OR ``Transgender'' OR ``Trans'' OR ``LGBT''\\OR ``Queer'' OR ``non*binary'')~And (``tech'' OR ``developer'')\\site:dev.to} & LGBT tech \\
{Filtered Documents:\\ D1, D2, D3, D4, D5,~D6,~D7, D8, D9} & 
{Filtered Documents:\\E1, E2, E3, E4, E5,~E6}\\
\end{tblr}
\label{string_grey}
\end{table}

\noindent\textbf{Inclusion and Exclusion Criteria:} we defined the following criteria based on Garousi et al. \cite{garousi2019guidelines}: (A) Materials accessible on the Web or by contacting the authors - to allow being able to trace findings from the original study to each individual source; 
(B) Material available in text; (C) Material that answers at least one auxiliary question, partially or fully - to consider that material is linked to research question discussed, and; (D) Non-duplicated material. We also analyzed whether the author has expertise in the field by reviewing their LinkedIn profile. After filtering the documents using the criteria, they were read in full. In Strategy A, nine documents (posts) were included, in Strategy B, six documents were included. The documents included after classification were placed in a spreadsheet and assigned an ID. For documents in strategy A, the prefix "D" was added, followed by the order of discovery. For documents in strategy B, the prefix "E" was added, followed by the order of discovery (Table \ref{string_grey}).

\noindent\textbf{Thematic synthesis from grey literature:} We applied the same process used in SLR (Section A). We obtained 23 themes\footnote{https://figshare.com/s/f491b3e9d61f65e02898 . Due to a recommendation from the literature \cite{santos2024patterns, mollen2024towards}, we chose to submit the grey literature synthesis data only for review and to provide only the codebook in the final version. These data obtained from public social networks will not be attached to the final publication.}. The main themes are discussed in the next section.

\subsection{GLR Results}

\begin{tcolorbox}[colback=Thistle!10!white,colframe=Thistle!75!black,title=A.1: \rqAUm]
\footnotesize
\textit{Professionals express frustration and dissatisfaction with their team dynamics. They face challenges such as difficulties in sharing identities, feelings of isolation, a lack of perceived diversity within the team, and the prevalence of male dominance. Despite these issues, they find satisfaction in participating in hackathons and collaborating with teams that prioritize diversity.}
\end{tcolorbox}

In \textbf{D1}, a transgender professional in a software team describes her experience at a start-up. She received \textbf{\faComments[regular] positive treatment when sharing gender identity} but was denied a promotion despite wanting a reassignment for six months. She emphasizes her dedication to resolving technical issues, including coding and training, and addresses key points relevant to the technology field. \emph{``Just because I don't have an engineering degree doesn't mean I don't know anything about technology or that I can't easily understand it (...)''} 

In \textbf{D2}, a transgender professional recounts aspects of her pre-transition experience that conformed to the dominant masculinity stereotype in the technology industry. She describes the weight \textbf{\faComments[regular] of being a transgender woman in technology}
\emph{``I've been on the other side of that (...) Before I transitioned, I was definitely bathing in the toxic masculinity that constitutes the tech industry. I spoke loudly, I was impetuous, I would definitely interrupt when others were speaking... here's the thing: it was effective.''}

The professional brings her considerations the \textbf{\faComments[regular] implicit male code in technology} :\emph{``There is something present in today's technology culture that is indescribable, a code of how we act and how we discuss that is implicit (...)} and also the \textbf{\faComments[regular] low perceived diversity} \emph{`` (...) Most of the time, it is just cis men talking to cis men, with no diverse voices in the room to change the pattern of conversation.''}.  In \textbf{D5}, the professional describes the evolution of her professional journey, which goes from frustration with perceived diversity to the pressure of being a minority and having a \textbf{\faComments[regular]} \textbf{ weight of representation}: \emph{``I felt the weight of representing multiple marginalized groups. So I leaned forward, I put my hands on my hips when I addressed the room, I spoke loudly, got serious, smiled less (...)''}. The professional also highlights the positive impact of \textbf{\faComments[regular] a supportive and inclusive team environment} \emph{``My team was supportive and kind; they guided me through the tech jargon and answered my silly questions with patience and respect}. And underscores the significance of \textbf{\faComments[regular] recognizing and valuing effort} and \textbf{\faComments[regular] a learner’s mindset}:  \emph{``And most of all, they appreciated my efforts to understand their roles...I made up for in genuine curiosity and a willingness to learn.''}

\textbf{D3} emphasizes the importance of intentionally including individuals from underrepresented gender identities to  \textbf{\faComments[regular] increase diversity}. The specific mention of trans women, non-binary, and intersex people underscores the need to expand beyond traditional diversity categories: \emph{``Making sure to include trans women, non-binary and intersex people in the group goes a long way towards inclusivity.''}. In \textbf{E2}, the developer shares that her introduction to the tech industry began at a \emph{bootcamp} focused on LGBTQIA+ inclusion. She reports facing \textbf{\faComments[regular] challenges during her learning experience}, particularly about team dynamics. In \textbf{E4}, the developers talk about the challenges they faced in learning to code. The reports from the professional highlight the challenges transgender individuals face in agile software development teams, including \textbf{\faComments[regular] psychological insecurity}, \textbf{\faComments[regular] fear of prejudice}, and \textbf{\faComments[regular] concerns about professional growth, team dynamics, and engagement}.

In \textbf{E5}, a developer talks about feeling motivated to participate with others LGBTQIA+ in a \emph{hackathon} but also the uncertainty about whether their experience aligns with standard hackathon dynamics - \textbf{\faComments[regular] uncertainty about norms}. The person describes \textbf{\faComments[regular] positive outcomes through experience and networking}: \emph{``I'm not sure if this experience is typical of hackathons... The experience and networking I sought were great...''}


\begin{tcolorbox}[colback=RoyalPurple!10!white,colframe=RoyalPurple!75!,title=A.2: \rqADois]
\footnotesize
\textit{Professionals in the tech industry often face challenges such as toxicity, implicit and masculine codes of conduct, discrimination, and impostor syndrome. There are also expectations for greater diversity, more inclusive teams and job interviews, support for underrepresented minorities—including transgender and non-binary people—and the elimination of gender bias.}
\end{tcolorbox}


Themes associated with developers' expectations are marked with the symbol \faFlag[regular], while themes related to challenges are marked with \faFrown[regular].

In \textbf{D1}, the professional reports fear due to \textbf{\faFrown[regular] suffering discrimination} in the workplace: \emph{``Just because I don't have an engineering degree doesn't mean I don't know anything about technology (...) I'm scared to be in the room with you. (...) Being transgender in tech is even harder."}.  In \textbf{D7}, an episode of discrimination is reported. \emph{``I have suffered discrimination (...)"}. In \textbf{D1} and \textbf{D5}, \textbf{\faFrown[regular] impostor syndrome} is reported as a challenge in the professional journey.





In \textbf{D8,} the LGBTQIA+ developer's expectations involve positive expectations regarding \textbf{\faFlag[regular]  support for LGBTQIA+}. In \textbf{D9}, the developer reports positive expectations around the \textbf{\faFlag[regular] expansion of gender in technology movements}, which includes being a non-binary person.

In \textbf{E1,} the developer (\emph{Queer)} reports that her journey in the technology area is lonely and faces several challenges, despite reporting support from members within the team. There are \textbf{\faFrown[regular] social isolation and exclusion} \emph{``Still, I feel lonely in the 
tech scene as an openly queer man (...)}, \textbf{\faFrown[regular] contrast between negative experiences and support} \emph{``I tend to get a front row seat to the sexism, anti-queerness, and toxic masculinity that tech spaces have gotten a bad rap for (...) My colleagues were always there to support me...''}

In \textbf{E3}, a transgender and non-binary developer shares their transition and challenges in the tech industry, addressing \textbf{\faFrown[regular] discomfort with incorrect pronouns}: \emph{``Listen to us and what we have to say. Believe us when we tell you about our experiences. Make space for us. Especially for non-binary people (...)''}. In \textbf{E4}, reported expectations include \textbf{\faFlag[regular] more diverse environments}, inclusion of non-binary and transgender individuals, \textbf{\faFlag[regular] diversity-focused initiatives}, \textbf{\faFlag[regular] career growth opportunities} in software, and greater \textbf{\faFlag[regular] acceptance of diverse gender identities}.

Among the challenges, the inclusion of transgender developers found in the thematic synthesis resonates with data from the literature \cite{prado2020trans, ford2019remote,  de2023benefits}. This underrepresentation in the software industry is challenging for agile teams, as these professionals routinely face transphobia, isolation, and psychological insecurity,  affecting retention in teams\cite{prado2020trans, de2023benefits, de2023transgender}.

\begin{tcolorbox}[colback=ForestGreen!10!white,colframe=ForestGreen!75!black,title=A.3: \rqATres]
\footnotesize
\textit{It’s essential to include women, transgender, non-binary, and intersex individuals in professional settings. Efforts should focus on promoting diversity and raising awareness to reduce bias. This involves creating space for LGBTQIA+ professionals, eliminating assumptions about women, addressing unconscious biases, and challenging the belief that leaders must be men.}
\end{tcolorbox}

\textbf{\textit{Considering the DX dimensions and the documents  we derived the following recommendations:}}
Regarding \textbf{conation}, \textbf{D3} emphasizes the inclusion of trans women, non-binary, and intersex individuals as a step toward actively embracing equality in the tech industry. Encourage companies and teams to include underrepresented groups in their initiatives actively. \textbf{D7} prioritizes awareness and educational efforts in the workplace to reduce unfounded fears and prejudices against underrepresented groups in technology. Promote structured training and workshops to address biases and foster inclusivity. Finally, \textbf{E6} addresses unconscious biases in hiring practices, such as assumptions about women’s technical capabilities, to lower artificial barriers for underrepresented groups in the technology sector. Create actionable plans to eliminate biases in executive and senior leadership expectations.

Concerning \textbf{cognition}, \textbf{D6} acknowledges the impact of non-inclusive language in team settings. If a term or phrase makes someone feel excluded, replace it with a more inclusive alternative. Foster an environment where language evolves to accommodate all team members. \textbf{D7}, highlight the significance of awareness-raising actions and educational work to combat prejudice. Encourage team discussions and learning sessions to address the root causes of workplace biases. \textbf{E4} recommends focusing on creating space and genuinely listening to LGBTQIA+ individuals, especially non-binary and transgender professionals, to better understand their experiences and needs in the technology industry.

About \textbf{affect}, \textbf{D4} recommends creating an atmosphere where all individuals feel genuinely welcome, rather than targeting specific proportions of ``diverse" participants. Prioritize inclusion as a cultural value that transcends numerical representation.
When organizing events, as discussed in \textbf{D4}, aim to include all programmers without fragmenting into overly specific groups, while ensuring that diverse identities feel represented and supported. Then, \textbf{D6} addresses the emotional toll of microaggressions faced by underrepresented groups, such as being undervalued despite expertise. Encourage teams to support actively and uplift colleagues who experience these subtle but harmful behaviors.

\section{Discussion and Practical Actions}

\textbf{The triangulation of (A.1) findings show that} diverse teams demonstrate higher productivity and resilience due to the unique perspectives of their members. SLR studies, such as S1 and S10, highlight these benefits, corroborated by GLR documents like D5 and E5, which emphasize the positive impact of inclusive team environments and events, such as hackathons, on LGBTQIA+ professionals' collaboration.

Psychological safety and identity management emerged as crucial factors for LGBTQIA+ professionals \textbf{\textit{at the individual level}}. SLR findings (S2) underline the advantages of remote work, which allows individuals to control the disclosure of their identity, enhancing their integration into teams. However, persistent challenges such as invisibility and isolation were noted. Similarly, GLR accounts (D1, E2, E3) reflect these dynamics, indicating that while remote work supports identity management, it does not address the emotional toll of isolation. 

\textbf{\textit{At the team level}}, implicit cultural codes and toxic masculinity in agile environments present significant barriers to inclusion. Studies (S7, S8) and GLR sources (D2) report that these norms stifle diverse conversations and create challenges for LGBTQIA+ professionals. Addressing these cultural issues requires targeted interventions, such as conducting regular diversity and inclusion (D\&I) workshops to dismantle toxic behaviors and foster allyship. Introducing team rituals and norms that encourage inclusive communication and collaborative problem-solving can also promote a more welcoming environment. Furthermore, implementing visibility measures, such as recognizing LGBTQIA+ contributions in team discussions and events, enhances the sense of belonging.

Visibility and diversity are crucial for inclusive team dynamics. S1 highlights issues like poor communication channels, while GLR (D3, E4) stress the need for active representation, especially for transgender and non-binary professionals. Teams can address this by implementing structured onboarding to introduce D\&I policies and resources. Regular events, like inclusive hackathons, reinforce inclusion and build community.

\textbf{\textit{At the organizational level}}, systemic barriers such as invisibility and isolation persist, particularly in remote settings. Both SLR (S2) and GLR (D1, E2) findings underscore the need for proactive strategies to improve team integration and engagement. Organizations should develop and enforce comprehensive D\&I policies that address specific challenges faced by LGBTQIA+ professionals. Investing in leadership training to cultivate inclusive managers who actively promote psychological safety and visibility within their teams is equally important. Finally, fostering partnerships with LGBTQIA+ advocacy organizations enhances representation and provides access to external support networks, reinforcing the organization’s commitment to diversity and inclusion.

\textbf{Concerning (A.2),} LGBTQIA+ professionals often expect inclusive environments with effective D\&I policies, representation, and psychological safety. S1, S6, S10 highlight the value of team acceptance and inclusive leadership, while GLR (D3, E4) emphasize equitable treatment, especially for non-binary and transgender individuals. However, recurring challenges include discrimination, microaggressions, and the toll of identity management. SLR (S2, S4, S7, S8) report bias in team interactions, weak D\&I initiatives, and the emotional burden of representing multiple marginalized identities. GLR (D1, D7, E1, E3) echo these, noting exclusion, toxic behaviors, and discomfort with misused pronouns, which foster isolation and hinder growth. \textbf{\textit{At the individual level}}, organizations can implement strategies to mitigate these barriers. Providing mentorship programs that help LGBTQIA+ professionals navigate workplace challenges and foster a sense of belonging is critical. Additionally, offering resources such as LGBTQIA+ employee networks, support groups, and access to inclusive mental health services can promote psychological well-being. Encouraging tailored career development plans that recognize individual LGBTQIA+ experiences and aspirations.

\textbf{\textit{At the team level}}, fostering inclusivity is key to overcoming systemic barriers. SLR findings (S2, S10) highlight the need to give LGBTQIA+ professionals control over identity disclosure, while GLR accounts (D5, E3) stress the tension between authenticity and self-disclosure risks. Impostor syndrome, especially for transgender and non-binary developers, adds further challenges (SLR S6, GLR D1, D5). Teams can address these by conducting D\&I workshops on allyship and inclusive communication, celebrating diversity with team rituals, and ensuring structured onboarding covers D\&I policies and pronoun usage to support new members.

Despite challenges, remote and hybrid work models help mitigate negative experiences. SLR (S2, S11) discuss the safety and flexibility they provide, while GLR sources (D3, E4) emphasize greater autonomy and protection from harmful interactions. \textbf{\textit{At the organizational level,}} systemic change is essential to foster inclusivity. Key steps include developing robust D\&I policies addressing gender identity and expression, training leaders to create safe spaces for identity-related discussions, and partnering with LGBTQIA+ advocacy groups to enhance representation and access to resources. These actions reinforce the organization's commitment to inclusivity.

\textbf{When discussing (A.3), }recommendations can be divided into short-term and long-term strategies. \textit{\textbf{In the short-term,}} organizations should prioritize offering remote and hybrid work options, enabling LGBTQIA+ employees to control their environments and disclosure preferences while mitigating risks such as microaggressions and discrimination (SLR: S2, S8, S10; GLR: D3). Transparent anti-discrimination policies with clear reporting protocols must be drafted and shared publicly to address harassment and bias effectively (SLR: S2, S7, S8; GLR: D4, E3). Training on inclusive language and pronoun usage should be implemented, alongside audits to remove exclusionary terminology (GLR: D3, D6). Diversity can be celebrated through initiatives like Pride Month events and storytelling sessions, fostering team cohesion (GLR: D5, E6). Establishing peer support groups or committees for LGBTQIA+ professionals further promotes safe spaces and reduces isolation (SLR: S2, S8, S10; GLR: D7).

\textit{\textbf{Long-term strategies should focus on systemic changes that embed inclusivity into the organization's culture.}} Recruitment and onboarding processes should emphasize transparency and inclusivity by explicitly mentioning diversity commitments and offering tailored mentorship for LGBTQIA+ employees (SLR: S8, S10; GLR: D3, E4). Leadership development initiatives should actively involve LGBTQIA+ individuals in decision-making roles while providing diversity training for all leaders to address unconscious biases and foster supportive team environments (SLR: S8; GLR: D3, E6). Inclusive event organizations, such as hackathons, should ensure participation from diverse groups and adopt codes of conduct to promote a welcoming atmosphere (SLR: S7, S8; GLR: E4). Comprehensive Diversity and Inclusion (D\&I) programs with measurable long-term goals should be established and regularly assessed for effectiveness (SLR: S7, S10; GLR: E6). Organizations should also integrate diversity into strategic goals as a driver for innovation and collaborate with external experts to implement best practices (SLR: S10; GLR: D3).

Invisibility and isolation can be mitigated by mentoring programs and inclusive team-building activities that enhance visibility and engagement (SLR: S2; GLR: D7). Toxic work environments require training for managers and employees to recognize and address microaggressions, foster accountability, and support affected individuals (SLR: S7, S8; GLR: D4). Policies and tools that empower LGBTQIA+ professionals to manage identity disclosure can reduce emotional and professional risks, ensuring psychological safety and autonomy (SLR: S2, S8; GLR: E3, E4). By implementing these strategies, organizations can create inclusive environments that improve the experiences of LGBTQIA+ professionals and leverage the innovation potential of diverse agile teams.

\section{ Threats to validity}

The limited number of relevant studies poses challenges in gathering evidence on underrepresented populations, such as non-binary and transgender individuals, within the LGBTQIA+ community. To address this, both SLR and GLR were conducted to enhance the reliability of the results. Combining these approaches ensured broader coverage of available and published information. Another threat to validity lies in the small amount of material identified based on the criteria for synthesizing grey literature. To mitigate this, three established recommendations from the literature \cite{kamei2021evidence}, \cite{garousi2019guidelines}, \cite{garousi2020benefitting} were followed. Additionally, two web search strategies, Strategy A and Strategy B (Section \ref{glrMethod}), were defined to gather comprehensive material. This careful process was crucial to maintaining the quality of the results. The integration of data from industry professionals may be influenced by personal biases linked to their positions and contexts. Unfortunately, this threat could not be mitigated.

\section{Conclusion and Future work}
This research focused on gathering evidence on the experiences of diverse sexual and gender identities within the community of industry professionals who identify as members of the LGBTQIA+ population. However, existing literature highlights the need to explore the intersections of this population with race, individuals with disabilities, and neurodivergent individuals. Future studies could collect data on these intersections and examine their implications for the experiences of LGBTQIA+ professionals.
\bibliographystyle{ieeetr}

\end{document}